\documentclass[aps.pra,twocolumn,showpacs,preprintnumbers,amsmath,amssymb]{revtex4}
\usepackage{mathrsfs}
\usepackage{graphicx}
\usepackage{amsmath}
\usepackage{ulem}
\usepackage{color}

\begin{document}

\title{Modeling dynamics of entangled physical systems with superconducting quantum computer}
\author{A. A. Zhukov$^{1,2}$, W. V. Pogosov$^{1,3}$, Yu. E. Lozovik$^{1,4,5}$}

\affiliation{$^1$Dukhov Research Institute of Automatics (VNIIA), 127055 Moscow, Russia}
\affiliation{$^2$National Research Nuclear University (MEPhI), 115409 Moscow, Russia}
\affiliation{$^3$Institute for Theoretical and Applied Electrodynamics, Russian Academy of
Sciences, 125412 Moscow, Russia}
\affiliation{$^4$Institute of Spectroscopy, Russian Academy of Sciences, 142190 Moscow region,
Troitsk, Russia}
\affiliation{$^5$Moscow Institute of Physics and Technology, 141700 Moscow region, Dolgoprudny, Russia}

\begin{abstract}
We implement several quantum algorithms in real five-qubit superconducting quantum processor IBMqx4 to perform quantum computation of the dynamics of spin-1/2 particles interacting directly and indirectly through the boson field. Particularly, we focus on effects arising due to the presence of entanglement in the initial state of the system. The dynamics is implemented in a digital way using Trotter expansion of evolution operator. Our results demonstrate that dynamics in our modeling based on real device is governed by quantum interference effects being highly sensitive to phase parameters of the initial state. We also discuss limitations of our approach due to the device imperfection as well as possible scaling towards larger systems.
\end{abstract}


\author{}
\maketitle
\date{\today }

\section{Introduction}

Quantum computers and simulators are prospective for the resolution of problems which can hardly be solved using conventional computing systems. In principle, these quantum devices can be constructed on the basis of different physical platforms, see, e.g., Refs. \cite{Girvin,Blatt,Lukin,Harty,Lu,Maller,Zwan,Nori}. However, over last years, superconducting realization seemed to become most promising for the construction of programmable quantum computers. Quantum processors of several types have been created and various algorithms have been implemented to show concepts of error correction \cite{Matrinis1,DiCarlo,Gambetta,Chow}, modeling spectra of molecules \cite{variat} and other fermionic systems \cite{Hubbard}, simulation of light-matter systems \cite{LM}, many-body localization \cite{MBL}, machine learning \cite{ML}, scaling issues \cite{Rigetti} etc. Besides, superconducting quantum circuits provide a unique platform to study the effects of quantum optics and nonstationary quantum electrodynamics, see, e.g., Refs. \cite{Devoret,Astafiev,Macha,Oelsner,DCE1,DCE2,Segev,We1,We2,We3}.

State-of-the-art processors contain up to sixteen superconducting qubits with individual control and readout. It is highly possible that a similar device containing fifty qubits will appear in the very near future \cite{supremacy1,Chow}. Quantum computer of such a size might enable for the first time to demonstrate "quantum supremacy" over modern and most capacitive conventional supercomputers \cite{supremacy1}. The idea of this crucial demonstration is to create a highly entangled state of nearly fifty qubits, the dimension of Hilbert space needed to store it being $2^{50}$. Storage and manipulation of such a state is beyond the capabilities of best modern supercomputers. Despite of the impressive progress, this task is very difficult in the view of decoherence processes inevitable in real quantum devices.

Several superconducting quantum processors, created within IBM Q project, are available through the internet via IBM cloud. In the present paper, we report on experiments with one of such processors -- five-qubit IBMqx4 chip. This chip is used by us to study dynamics of interacting spin-1/2 particles starting from different initial conditions. We focus on effects arising due to the presence of  entanglement of several particles in the initial state - two-particle entangled state and three-particle entangled state, the latter being a quantum superposition of a couple of two-particle entangled states. We show how degrees of freedom of modeled system can be identified with physical qubits of the chip, despite of limitations of its topology, and encode these initial states in the device. The free evolution of the system is implemented in a digital way using one-step Trotter expansion of the evolution operator. We demonstrate high sensitivity of the dynamics to the entanglement and phase factors of the initial state. Thus, we show that our modeling based on real device is governed by quantum interference effects. Experimental results are in a good semi-quantitative agreement with theoretical expectations.

However, attempts to go beyond one-step Trotterrization lead to the dramatic suppression of quantum interference effects in our modeling, which elucidates limitations of current technology and stresses difficulties in demonstration "quantum supremacy" using entangling operations between many qubits. The main sources of imperfections are errors of two-qubit gates as well as decoherence processes. Nevertheless, we believe that further technological improvements as well as scaling towards computers with tens of qubits will indeed allow to reach a modeling the dynamics of quantum systems difficult to study with conventional supercomputers.

Our paper is organized as follows. In Section II we describe the architecture of IBMqx4 chip and discuss physical systems most suitable for quantum computation with this chip. We present Hamiltonians of these systems and show how degrees of freedom of modeled systems can be mapped on degrees of freedom of the physical device. In Section III, we discuss various initial conditions, which include both entangled and disentangled states, and their encoding in the device. In Section IV, we explain how we implement dynamics using one-step Trotter expansion of evolution operator. Section V presents experimental results of our modeling with quantum computer on the dynamics of modeled system. We also provide a comparison with theoretical results based on the same one-step Trotter expansion, but for the ideal system, i.e., in absence of decoherence, as well as gate and readout errors. Section VI summarizes our results and conclusions.

\section{Mapping physical system on physical chip}

\begin{figure}[h]
\includegraphics[width=0.45\linewidth]{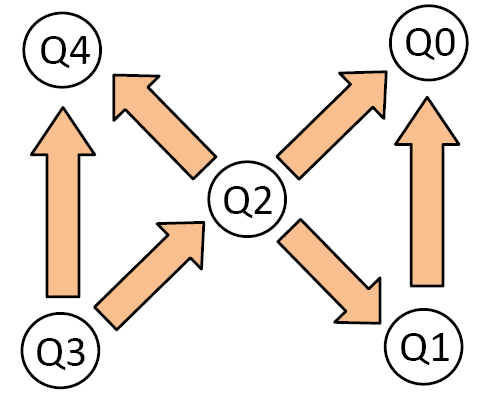}
\caption{
\label{chip}
 (Color online) Schematic view of IBMqx4 chip. Two-qubit gates and their directions are shown by arrows (see in the text).}
\end{figure}

The structure of superconducting quantum processor IBMqx4 is shown schematically in Fig. 1. The central qubit Q2 is connected by CNOTs to four remaining qubits Q0, Q1, Q3, and Q4, as indicated in Fig. 1 by arrows. Each arrow points from the control qubit to the target qubit.

The most evident approach is to associate this architecture with a central spin model, which describes an ensemble of spin-1/2 particles, the central particle interacting with all other particles (bath), see, e.g., Ref. \cite{Prokofiev}. The interaction between the central particle and particles of the bath can be implemented in a digital way using CNOTs, which connect central qubit with four others (for details on implementation of interaction, see Section IV). The particles of the bath either do not interact with each other directly or do interact: additional CNOTs between Q3 and Q4 as well as between Q0 and Q1 can be used to implement digitally pairwise interaction of corresponding particles, whose quantum states are encoded into these four qubits. CNOTs between any two qubits can be also used to construct entangled quantum states of these two qubits. Thus, entangled initial states of modeled system can be directly mapped to the entangled states of physical qubits. This scheme suggests one-to-one correspondence between the states of modeled system of particles and physical qubits of the device, as dictated by interaction term of modeled Hamiltonian as well as the structure of the initial state of modeled system. However, in practice, we use several different schemes, which are described in the next Section. Particularly, in order to model the dynamics of the system with three entangled particles, we have to go beyond simple one-to-one mapping due to certain limitation of real chip topology.

We are going to analyze the dynamics starting from different initial conditions which include excited central spin and unexcited spins of the bath as well as more unusual condition: unexcited central spin and entangled quantum bath.

Note that central spin models are used to study decoherence processes in quantum dots and other systems. In physically realistic systems, particles of the bath have their own environments. However, influence of these additional baths can be neglected within some initial time interval during the free evolution, provided the coupling to the central spin is larger.

The simplest relevant model from the class of central spin models is XX central spin model with the Hamiltonian of the form
\begin{eqnarray}
H_{cs}=  \sum_{j=1}^L  \epsilon_j (\sigma_{j,z}+1/2)+ \epsilon_c (\sigma_{c,z}+1/2) + \nonumber \\
g \sum_{j=1}^L (\sigma_{c}^+ \sigma_{j}^- + \sigma_{c}^- \sigma_{j}^+),
\label{Hamiltoniancentralspin}
\end{eqnarray}
where $\sigma_{j,z}$ and $\sigma_{j}^\pm$ are Pauli operators associated with particles of the bath, while $\sigma_{c,z}$ and $\sigma_{c}^\pm$ refer to the central spin; $\epsilon_j$ and $\epsilon_c$ are excitation energies of spins of the bath, which do not interact with each other directly, whereas $g$ is the interaction constant between the central spin and each spin of the bath. For simplicity, we hereafter assume that all excitation energies are the same and switch to the rotating frame. The Hamiltonian we model ultimately reads as
\begin{eqnarray}
H =  g \sum_{j=1}^L (\sigma_{c}^+ \sigma_{j}^- + \sigma_{c}^- \sigma_{j}^+).
\label{Hamiltoniancentralspin1}
\end{eqnarray}

It is easy to realize that the same Hamiltonian is directly applicable also to spin-boson coupled systems. Physical realizations range from quantum optical systems \cite{Garraway} to Fermi-Bose condensates near the Feshbach resonance \cite{Andreev}. The most evident realization is an ensemble of $L$ two-level systems (spin-1/2 particles) coupled to a single mode electromagnetic field. The exact mapping to (\ref{Hamiltoniancentralspin1}) exists for Hamiltonians from this class, which conserve excitation number, and in the sector of a single excitation. Note that the nature of the two-level systems can be very diverse ranging from natural atoms to macroscopic artificial quantum systems. Suitable quantum optical system is described by Dicke Hamiltonian of the form
\begin{eqnarray}
H_{qo}=\epsilon \sum_{j=1}^L  (\sigma_{j,z}+1/2) + \omega a^{\dagger } a + \nonumber \\
 g \sum_{j=1}^L (a^{\dagger }+ a)(\sigma_{j}^- + \sigma_{j}^+),
\label{Hamiltonian0}
\end{eqnarray}
where $\epsilon$ is spin excitation energy, $\omega$ is the energy of the boson, $g$ is the coupling energy between spin and boson subsystems (dipole-dipole interaction); $a^{\dagger }$ and $a$ are bosonic creation and destruction operators, while $\sigma_{j,z}$, $\sigma_{j}^\pm$, and $\sigma_{j}^z$ are spin Pauli operators. If spin excitation energy is in a resonance with boson energy $\omega$, rotating wave approximation can be utilized, which neglects counterrotating terms in the Hamiltonian (\ref{Hamiltonian0}) of the form $g(a \sigma_{j}^- + a^{\dagger } \sigma_{j}^+)$. These omitted terms do not conserve excitation number, i.e., the number of bosons plus the number of excited atoms. The resulting excitation-number conserving Hamiltonian in the rotating frame can be represented as
\begin{eqnarray}
H_{qo}^{RWA}=g \sum_{j=1}^L (a^{\dagger } \sigma_{j}^-+ a \sigma_{j}^+).
\label{Hamiltonian1}
\end{eqnarray}
Let us restrict ourselves to the situation, when excitation number does not exceed 1. Within the relevant subspace of the whole Hilbert space, $a^{\dagger }$ and $a$ can be replaced by Pauli operators acting in the space of two allowed states of boson subsystem. Thus we replace boson by an additional two-level system and arrive at the Hamiltonian of XX central spin model (\ref{Hamiltoniancentralspin1}).

In order to bring the Hamiltonian (\ref{Hamiltoniancentralspin1}) to the form conventional in the field of quantum computation we rewrite operators $\sigma^+$ and $\sigma^-$ through operators $\sigma^{x}$ and $\sigma^{y}$. After simple algebra, we represent (\ref{Hamiltoniancentralspin1}) in an equivalent form as
\begin{eqnarray}
H = \frac{g}{2} \sum_{j=1}^L (\sigma_{c}^{x} \sigma_{j}^{x} + \sigma_{c}^{y} \sigma_{j}^{y}).
\label{Hamiltonian}
\end{eqnarray}

\section{Initiating the physical system}

We are going to address three different realizations of spin systems each being characterized by its own initial condition. In this Section we describe these three realizations and we also explain how the initial conditions can be encoded into the system of physical qubits in IBMqx4 quantum processor.

\subsection{Two entangled spin-1/2 particles}

Let us consider the system of two particles of the bath coupled to the central particle. We assume that the initial state of the whole system is an unexcited central spin and entangled particles of the bath. This state is given by
\begin{eqnarray}
\Psi (0) = |\downarrow\rangle \otimes \frac{1}{\sqrt{2}} \left(|\downarrow \uparrow\rangle + e^{i\varphi}|\uparrow \downarrow\rangle\right),
\label{init1}
\end{eqnarray}
which is parameterized by a single parameter $\varphi$. We further refer the two-particle entangled state appearing in Eq. (\ref{init1}) to as 2PES.

\begin{figure}[h]
\includegraphics[width=0.45\linewidth]{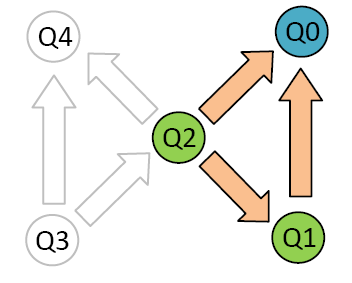}
\caption{ (Color online) Mapping between the modeled system of two spin-1/2 particles coupled to the central spin and elements of the physical device. Green circles denote physical qubits Q1 and Q2 which encode quantum states of the two particles, while blue circle denote physical qubit Q0 used to encode central spin state. Unused qubits and CNOTs are shown in grey.
\label{algo1}
 }
\end{figure}

\begin{figure}[h]
\includegraphics[width=0.45\linewidth]{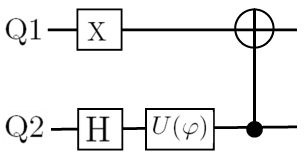}
\caption{
\label{2AES}
Quantum circuit for preparation of the two-particle entangled state 2PES (see in the text).}
\end{figure}

We associate the state of two particles with the quantum states of two physical qubits via one-to-one correspondence, as shown in Fig. \ref{algo1}. Q1 and Q2 are used to encode states of particles of the bath, whereas Q0 encodes quantum states of the central spin. CNOT between Q0 and Q1 as well as between Q0 and Q2 will be used to implement interaction between the central spin and the bath, as described in the next Section. CNOT between Q1 and Q2 is used to create an entangled state 2PES of these two qubits.

Note that multiple choices to perform mapping between the three spin-1/2 particles and physical qubits do exist for IBMqx4 chip even under restrictions originating from the form of interaction term of Hamiltonian (\ref{Hamiltonian}) and available CNOTs of the chip. We used the optimized one, which is based on minimization of total error induced by CNOT gates in our experiments, since different CNOTs of the chip exhibit different errors. The initial entangled state of two qubits can be prepared using the circuit shown in Fig. \ref{2AES}, where $U(\varphi)=\begin{bmatrix}
1 & 0 \\
0 & e^{i\varphi}%
\end{bmatrix}$: H and X are Hadamard and Pauli-X gates, respectively.

\subsection{Three entangled spin-1/2 particles}

\begin{figure}[h]
\includegraphics[width=0.95\linewidth]{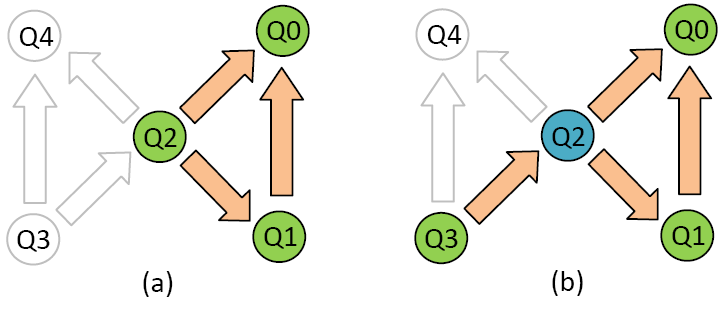}
\caption{
\label{algo2}
 (Color online) Mapping between the modeled system of three spin-1/2 particles coupled to the central spin and elements of the physical device. Green circles denote physical qubits which encode quantum states of three particles, while blue circle denote physical qubit used to encode central spin state. At the first stage (a) an entangled state of three physical qubits is created, which encode the entangled state of three particles via one-to-one correspondence. At the second stage (b) the state of the central qubit Q2 is transferred to the unused qubit Q3, while Q2 is further used to encode quantum state of the central spin.  Unused qubits and CNOTs are shown in grey.}
\end{figure}

\begin{figure}[h]
\includegraphics[width=0.6\linewidth]{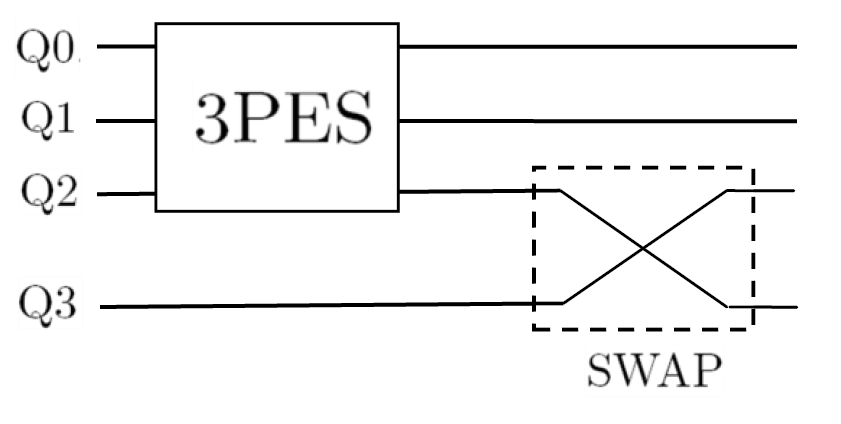}
\caption{
\label{3AES}
Quantum circuit for preparation of the three-particle entangled state 3PES encoded into the physical device.}
\end{figure}

We now consider the system of three spin-1/2 particles coupled to the central spin, the initial state of the system being
\begin{eqnarray}
\Psi (0) = |\downarrow\rangle \otimes \frac{1}{\sqrt{6}}\left(|\downarrow \downarrow \uparrow\rangle - 2e^{i\chi}|\downarrow \uparrow \downarrow \rangle + | \uparrow \downarrow \downarrow\rangle\right),
\label{init2}
\end{eqnarray}
where $\chi$ is phase parameter. The entangled state of three particles in Eq. (\ref{init2}) is further referred to as 3PES.

The mapping of this state to the physical system is less straightforward compared to the case of 2PES due to the limitations of the chip topology. The difficulty is in the fact that in order to create 3PES it is necessary to use three physical qubits with two CNOT gates between them. Therefore, central physical qubit Q2 of the chip has to be utilized. However, the same qubit has to encode quantum state of the central spin, since there must be three CNOT gates between it and three other qubits in order to model an interaction of the central spin and three particles of the bath. For this reason, the initial state (\ref{init2}) is prepared in two steps, as shown in Fig. \ref{algo2}. At the first stage, 3PES is created using Q0, Q1, and Q2 with the help of two CNOTs. At the second stage, the state of Q2 is transferred to Q3 using SWAP two-qubit gate, while Q2 is utilized to encode the state of the central spin. Thus, Q0, Q1, and Q3 are ultimately used to encode quantum state 3PES of three particles of the bath, whereas Q2 encodes quantum states of the central spin. The whole quantum circuit for preparation of the initial state is shown schematically in Fig. \ref{3AES}. The initial block used to prepare 3PES encoded into qubits Q0, Q1, and Q2 is described in Appendix A.

\subsection{Excited central spin and unexcited bath}

\begin{figure}[h]
\includegraphics[width=0.45\linewidth]{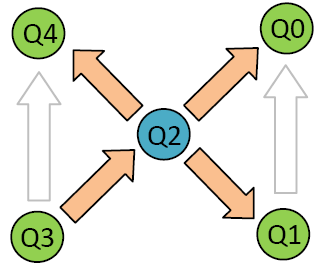}
\caption{ (Color online) Mapping between the modeled system of four spin-1/2 particles coupled to the central spin and elements of physical device. Green circles denote physical qubits which encode states of four particles, while blue circle denote physical qubit used to encode state of the central spin. Unused CNOTs are shown in grey.
\label{algo3}
 }
\end{figure}

Now we address the system of up to four spin-1/2 particles coupled to the central spin, the initial state of the bath being disentangled. We assume that bath initially contains unexcited particles, whereas central spin is excited
\begin{eqnarray}
\Psi (0) = |\uparrow \rangle \otimes |\downarrow \ldots \downarrow \rangle.
\label{init3}
\end{eqnarray}
Although it is straightforward to initialize this state, computations involving it enable us to utilize all the five qubits of the chip. Mapping between the modeled system and elements of the chip are shown in Fig. \ref{algo3}. Periphery qubits Q0, Q1, Q3, and Q4 are used to encode states of four particles of the bath, whereas Q2 encodes the state of the central spin.

Notice that the inversion of CNOT can be implemented in a standard way using additional Hadamard gates.

\section{Modeling dynamics via Trotterization}

\begin{figure}[h]
\includegraphics[width=0.95\linewidth]{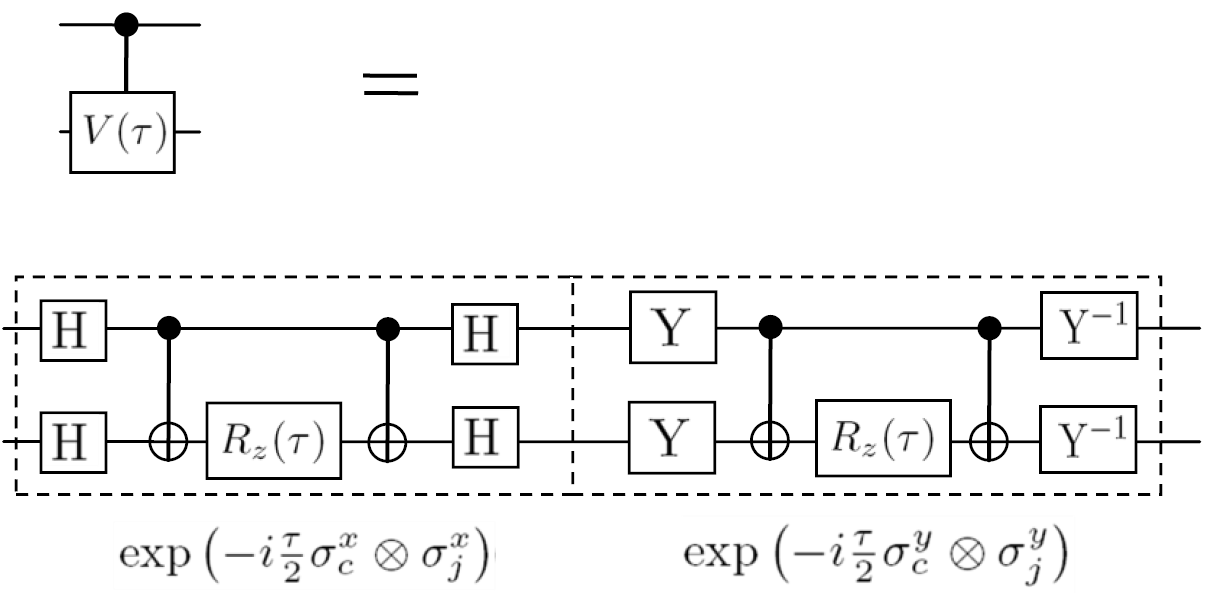}
\caption{
\label{interaction}
Quantum circuit for $\exp\left(-i\frac{\tau}{2} \sigma_{c}^{x} \otimes \sigma_{j}^{x}\right) \exp\left(-i\frac{\tau}{2} \sigma_{c}^{y} \otimes \sigma_{j}^{y}\right)$ (see in the text).}
\end{figure}

 We are now in the position to implement the dynamics of modeled system in a digital way using Trotterization. The free evolution of the system starting from the initial wave function $| \Psi (0) \rangle$ is given by the standard formula
\begin{eqnarray}
\Psi (t) = e^{-iHt}\Psi (0).
\label{evol}
\end{eqnarray}
Under one-step Trotter expansion, this expression is rewritten in the approximate form as
\begin{eqnarray}
\Psi (\tau) \approx \prod_{j=1}^{L} \exp\left(-i \frac{\tau}{2} \sigma_{c}^{x}\otimes \sigma_{j}^{x}\right) \exp\left(-i\frac{\tau}{2} \sigma_{c}^{y} \otimes \sigma_{j}^{y}\right)\Psi (0),
\label{evoltrot}
\end{eqnarray}
where dimensionless time $\tau=gt$ was introduced. Eq. (\ref{evoltrot}) is accurate at $\tau \ll 1$ ($t \ll 1/g$). $\sigma_{c}$ and $\sigma_{j}$ now refer to those physical qubits, which encode states of the central spin and particles of the bath, respectively, as described in the preceding Section.

Each gate the form $\exp\left(-i\frac{\tau}{2} \sigma_{c}^{x,y,z} \otimes \sigma_{j}^{x,y,z}\right)$ can be represented in a usual
 way thought the CNOT gate entangling two qubits as well as single-qubit gate
 $R_z(\tau)=\exp\left(-i\frac{\tau}{2} \sigma^{z} \right)$ in the appropriate basis.
 $R_z(\tau)$ is expressed through the standard IBMqx4 gate $U_3$ and Hadamard gate H as H$ U_3(\theta = \tau, \varphi = - \pi/2, \lambda = \pi/2)$H.
Fig. \ref{interaction} presents quantum circuit for $\exp\left(-i\frac{\tau}{2} \sigma_{c}^{x} \otimes \sigma_{j}^{x}\right) \exp\left(-i \frac{\tau}{2} \sigma_{c}^{y} \otimes \sigma_{j}^{y}\right)$, whereas Pauli-Y gate is expressed as $U_3(\theta = - \pi/2, \varphi = - \pi/2, \lambda = \pi/2)$.

By using a one-step Trotter expansion, we trace the time evolution of the system starting from three different initial conditions encoded into real physical device. In particular, we concentrate on the dynamics of mean population of the excited state of the central particle by measuring population of qubit, which encodes central spin states, at different values of dimensionless time $\tau$. The whole quantum circuits corresponding to three initial conditions are presented in Appendix B.

\section{Results and discussion}

\begin{figure}[h]
\includegraphics[width=1.00\linewidth]{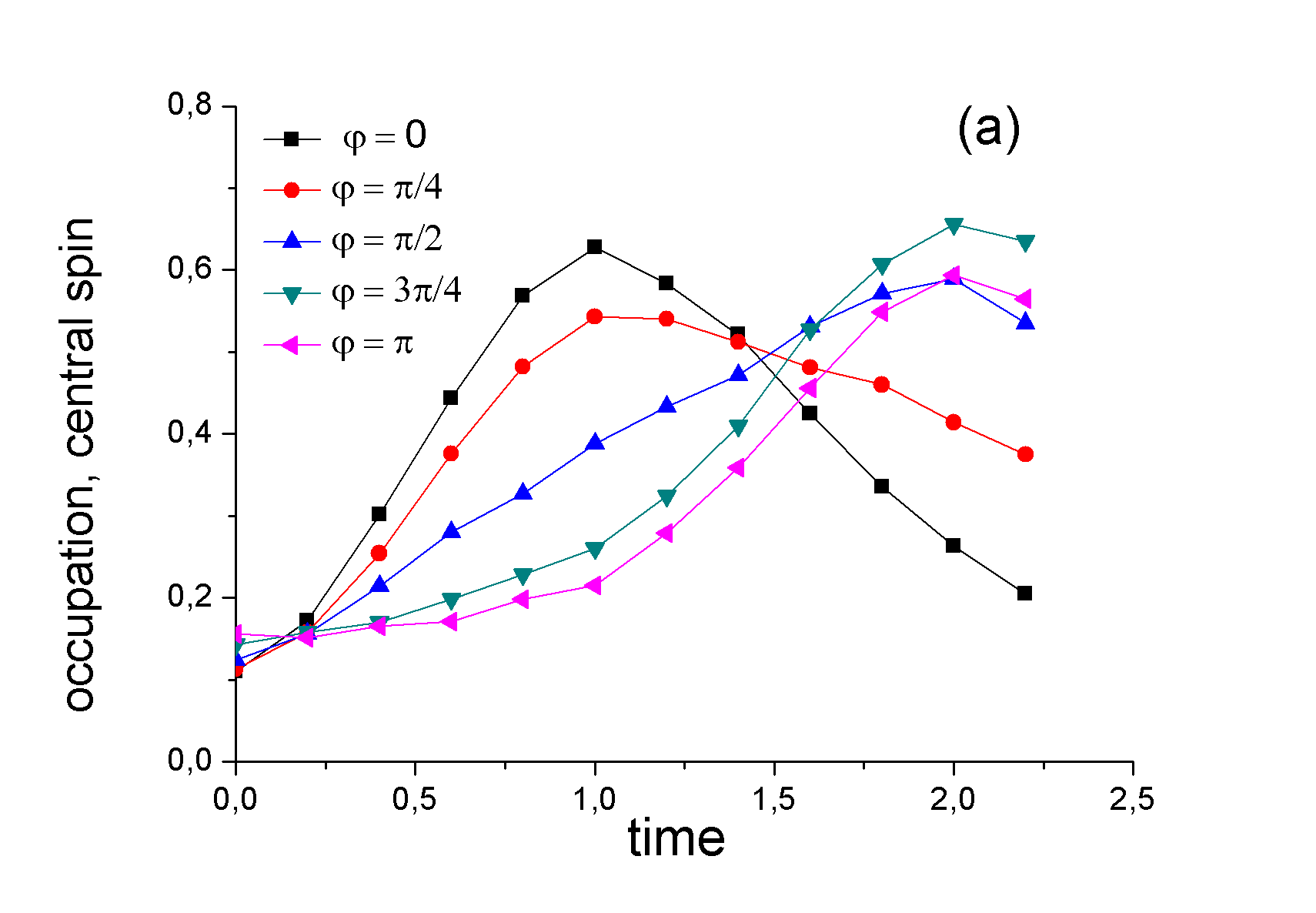}
\includegraphics[width=1.00\linewidth]{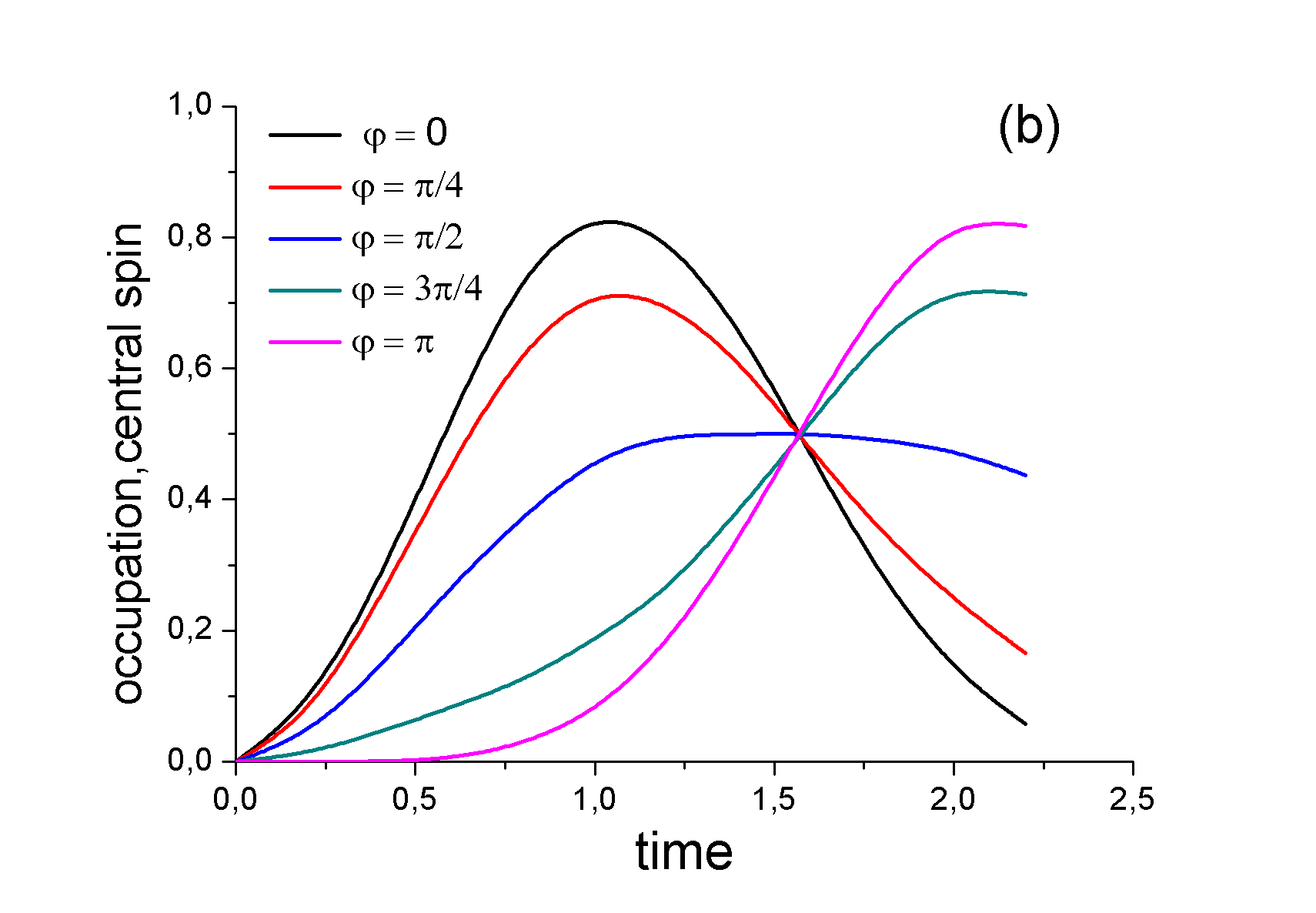}
\caption{
\label{results1}
 (Color online) The results of our experiment (a) and theory (b) for the mean population of the excited state of central particle as a function of time. The initial state of the system is two-particle entangled state of the bath and unexcited central spin. Different curves correspond to different values of phase parameter $\varphi$ entering the initial state. }
\end{figure}

Fig. \ref{results1} shows the evolution of the mean population of the excited state of central particle starting from the initial state (\ref{init1}) for several values of phase parameter $\varphi$. Fig. \ref{results1} (a) corresponds to the experimental results obtained with IBMqx4 quantum computer using quantum algorithms described in preceding Sections, while \ref{results1} (b) shows the results of theoretical predictions based on the same approximation -- one-step Trotter expansion of evolution operator. Both results should be identical in the case of an ideal quantum computer (no decoherence as well as gate and readout errors). The theoretical results can be readily found explicitly using Eq. (\ref{evoltrot}) or they can be obtained from IBM simulator (classical computer) available in IBM Q online system, which performs the same calculations numerically and can be used for the analysis of experimental results. We here work with data from this simulator because of its usability.

It is seen from Fig. \ref{results1} that there exists good semi-quantitative agreement between the results of experiment and theory. Dynamics is highly sensitive to the phase parameter $\varphi$; the dependencies on $\varphi$ are the same in Figures \ref{results1} (a) and (b). This sensitivity is unambiguous demonstration for the realization of entangled states and quantum interference effects in the device. These results are also of certain interest from the perspective of physics of the central spin model. They evidence that the excitation stored initially in the bath can be trapped in the bath without transferring to the central spin due to the entanglement and quantum interference effects.

Nevertheless, there exist significant deviations of experimental curves compared to theoretical ones. The reason is mainly in errors of CNOT gates which are typically several percent in IBMqx4 and thus induce quite large error after nearly ten CNOTs. Decoherence of physical qubits also gives noticeable contribution in the time scale of a single run of the algorithm. Of course, one-step Trotterization is also not exact, so it is accurate only at $\tau \ll 1$. A natural idea is to increase the number of Trotter steps. However, as we found experimentally, expanding the algorithm even to the two Trotter steps leads to almost completely smeared difference between curves for different values of $\varphi$, which in general shift downwards quite close to the $x$ axis. This illustrates certain limitations of state-of-art superconducting quantum processors.

The sensitivity of the population of the central spin to the phase parameter $\varphi$ can be understood by adopting quantum optics picture. Indeed, the value of phase parameter $\varphi$ equal to $\pi$ corresponds to the so called nonradiant or dark state, which is not coupled to light. This nonradiant state is an eigenstate of the Hamiltonian and therefore no mean photon occupation is induced upon the free evolution starting from this state (for the exact solution, i.e., inifine number of Trotter steps). The dynamics seen in Fig. \ref{results1} (b) for $\varphi = \pi$ at relatively large $\tau \sim 1$ is an artefact of one-step Trotter expansion. This expansion is reliable at $\tau \ll 1$, but it also provides adequate qualitative and even semi-quantitative results at $\tau \lesssim 1$. Tuning $\varphi$ from $\pi$ results in the increased coupling of the initial state to the light, as expected from the exact solution of the problem \cite{Garraway}. This behavior is reflected within the approximation based on one-step Trotter expansion: the initial growth of the mean photon number as a function of time becomes stronger when tuning $\varphi$ away from $\pi$ and reaches maximum at $\varphi=0$, see Fig. \ref{results1} (b). The behavior we expect from theory (one-step Trotter expansion) is reproduced in experiments, as seen in Fig. \ref{results1} (a).

\begin{figure}[h]
\includegraphics[width=1.00\linewidth]{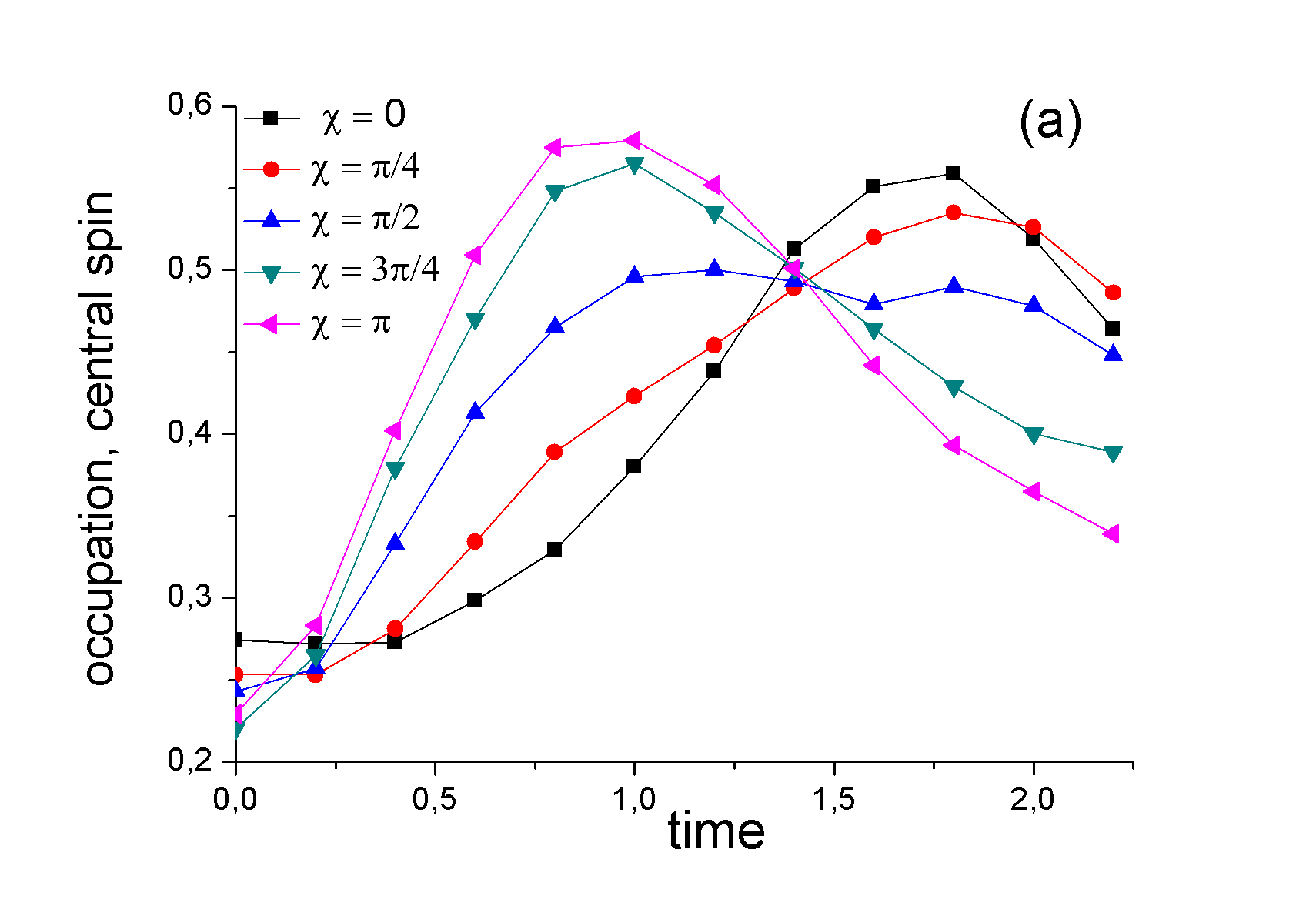}
\includegraphics[width=1.00\linewidth]{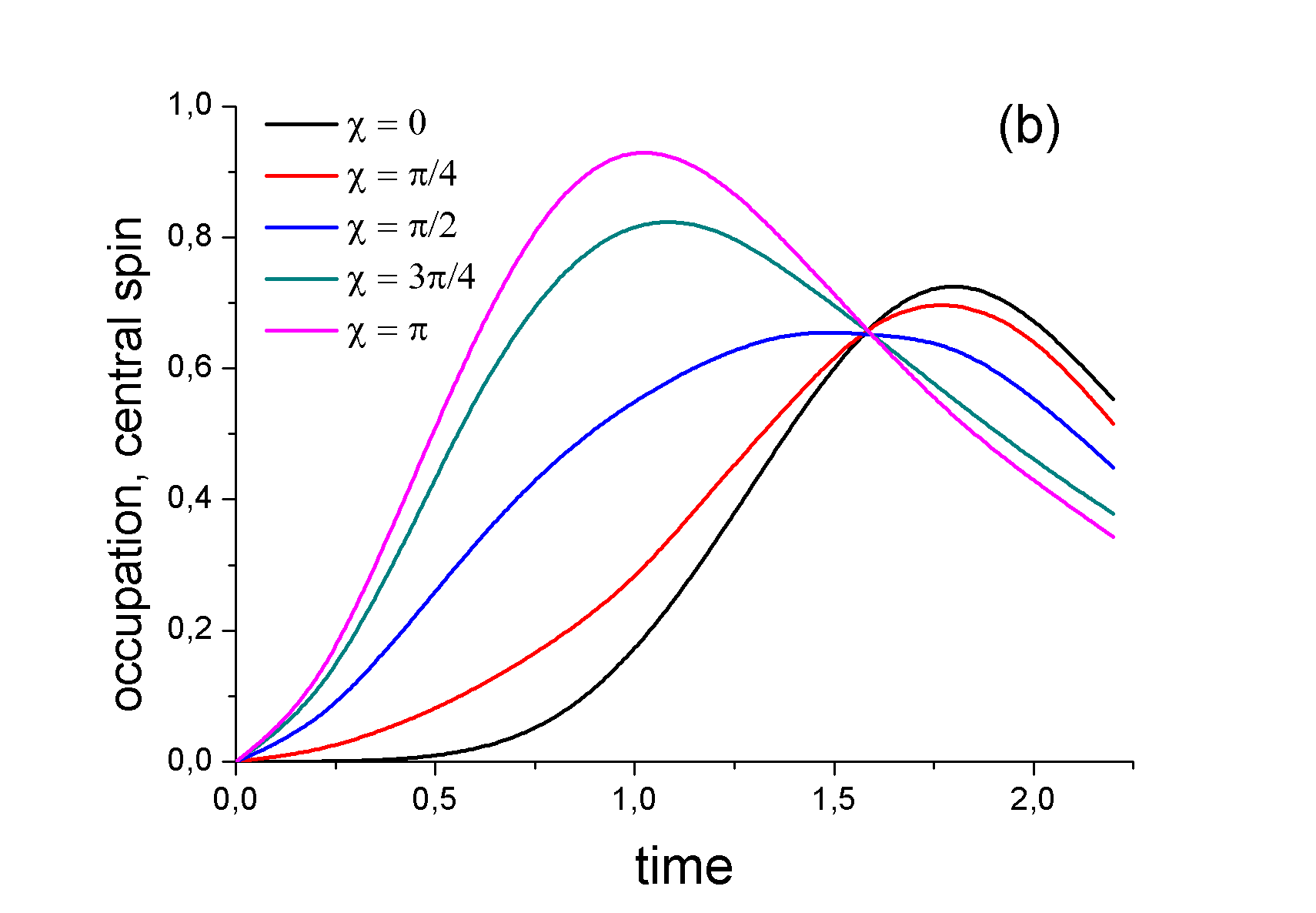}
\caption{
\label{results2}
 (Color online) The results of our experiment (a) and theory (b) for the mean population of the excited state of central particle as a function of time. The initial state of the system is three-particle entangled state of the bath and unexcited central spin. Different curves correspond to different values of phase parameter $\chi$ entering the initial state.  }
\end{figure}

Fig. \ref{results2} shows the evolution of the mean population of the excited state of central particle for the initial state (\ref{init2}) involving three entangled particles. There again exists a good semi-quantitative agreement between the theory and experiment. By theory we again mean an approximation based on one-step Trotter expansion. However, the sensitivity to phase parameter $\chi$ is less pronounced for experimental results (a) compared to the theoretical ones (b). The reason is in the increased length of the whole algorithm. Nevertheless, the dynamics is governed by quantum interference effects in this case as well. Particularly, the central spin is weakly occupied for $\chi = 0$ at $\tau$ small, i.e., for the initial three-particle entangled state of the form $|\downarrow \downarrow \uparrow\rangle - 2|\downarrow \uparrow \downarrow \rangle + | \uparrow \downarrow \downarrow\rangle$, which is a quantum superposition of two degenerate dark states $|\downarrow \downarrow \uparrow\rangle - |\downarrow \uparrow \downarrow \rangle$ and $| \uparrow \downarrow \downarrow\rangle - |\downarrow \uparrow \downarrow \rangle$.

\begin{figure}[h]
\includegraphics[width=1.00\linewidth]{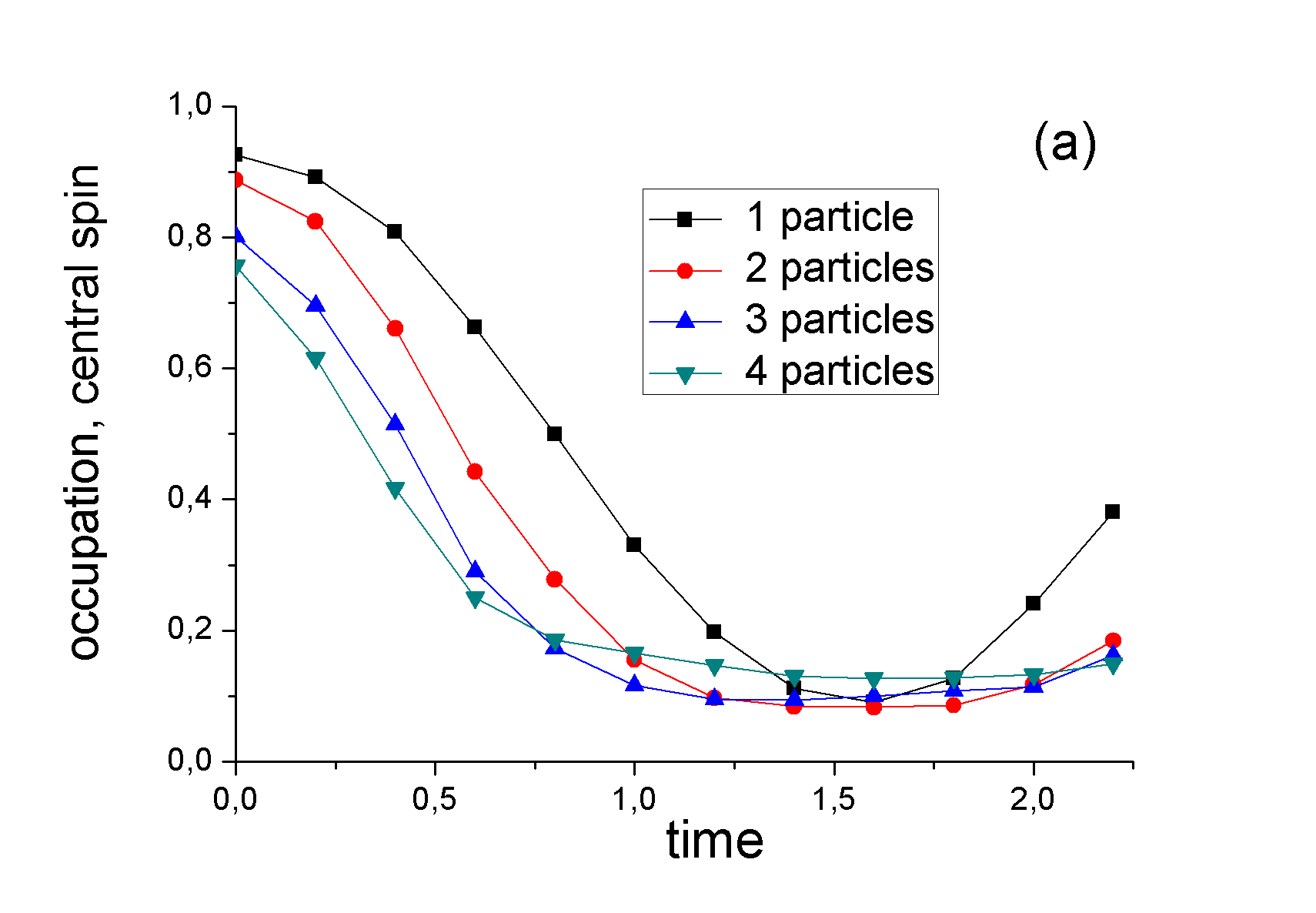}
\includegraphics[width=1.00\linewidth]{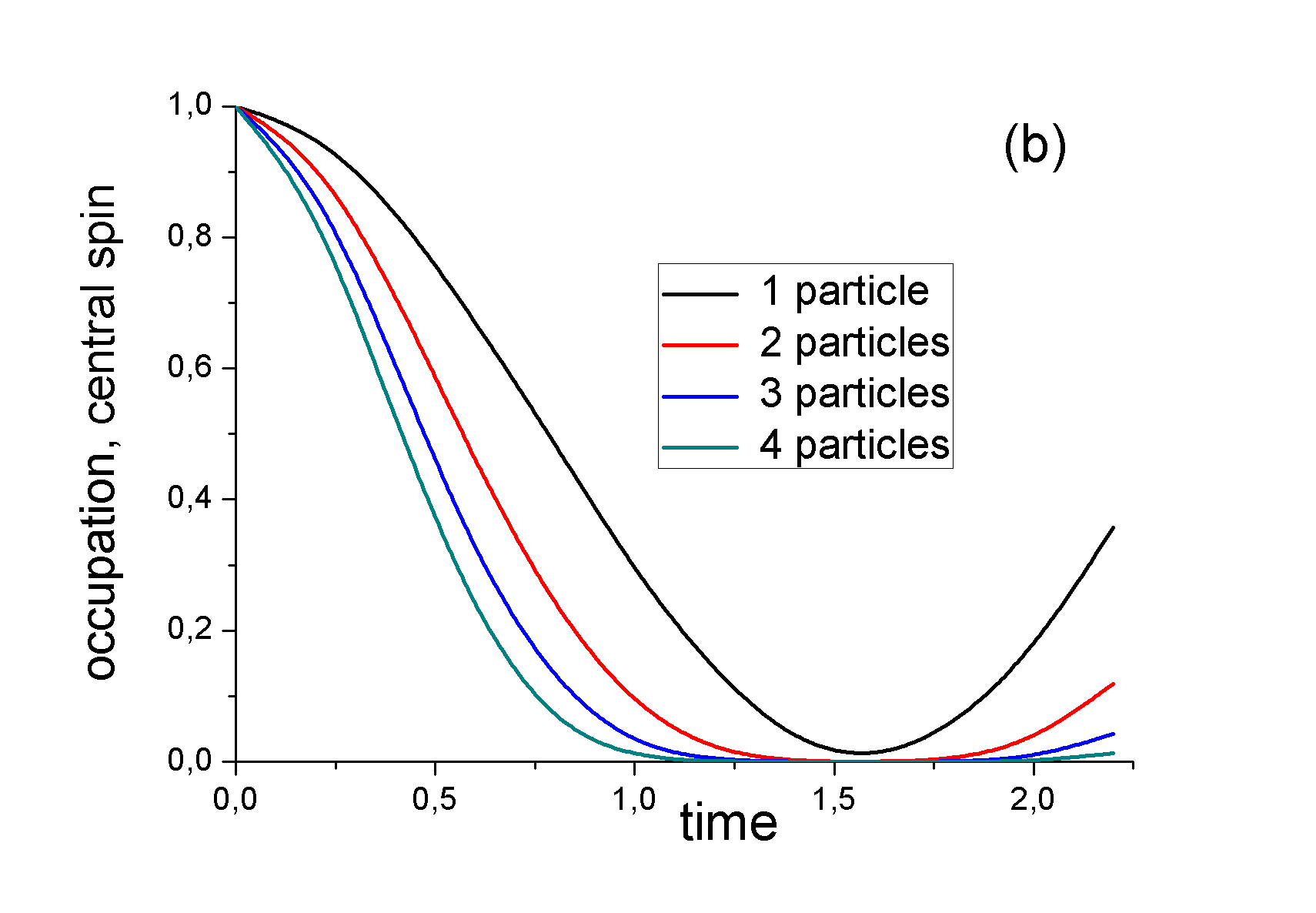}
\caption{
\label{results3}
 (Color online) The results of our experiment (a) and theory (b) for the mean population of the excited state of central particle as a function of time. The initial state of the system is excited central spin and $L$ unexcited spins of the bath. Different curves correspond to different values of $L$. }
\end{figure}

Let us now address dynamics starting from the initially disentangled bath (\ref{init3}): central spin excited and all spins of the bath unexcited. Time evolution of the population of the excited state of central particle for different numbers of spins of the bath, from 1 to 4, is shown in Fig. \ref{results3} using both the experimental (a) and theoretical (b) data (one-step Trotter expansion).

Adopting quantum optics understanding, within the exact solution, i.e., infinite number of Trotter steps, one would expect Rabi oscillations between the central spin and the collective spin constructed from individual spins of the bath. Thus, the initial excitation should freely transfer from central spin to the bath and back since no initial entanglement is present in the bath, which could block such a transfer. Moreover, a period of Rabi oscillations is expected to be proportional to $1/\sqrt{L}$, since the interaction energy between the central spin and the bath is enhanced as $g \sqrt{L}$. Fig. \ref{results3} (b) shows that such a collective behavior also exists under the one-step Trotter expansion. Comparing Fig. \ref{results3} (b) with the experimental results (a) obtained by quantum computer using the same one-step Trotter expansion embedded in the algorithm, we see that cooperative Rabi oscillations are indeed reproduced in experiments. Their period depends on the number of particles in a correct way being nearly proportional to $1/\sqrt{L}$.

\section{Summary and conclusions}

In the present paper, we implemented several quantum algorithms in the real five-qubit superconducting quantum computer IBMqx4. The aim of our work was to model the free evolution of small quantum systems starting from different initial conditions. The systems studied are spin-1/2 particles interacting either directly or through the boson field. They are described by central spin model and Dicke Hamiltonian, respectively.

We suggested a method to encode quantum states of spin-1/2 particles as well as boson into the quantum states of physical qubits of the chip taking into account limitations of chip's topology. We also showed how desired initial states, which include entangled or disentangled states of several particles, can be prepared in real chip. The entangled states addressed were two- and three-particles entangled states containing tunable phase factors. The three-particle state was composed from a superposition of a couple of two-particle entangled states.

The dynamics is addressed digitally using Trotter expansion of evolution operator. CNOT gates of the chip are utilized to construct both the entangled initial states and interactions between spins. Under a one-step Trotter expansion, we found good semi-quantitative agreement between the experimental data and theoretical predictions based on the same approximation. Particularly, the dynamics of the system in the experiment is shown to be highly sensitive to the entanglement of the initial state and phase parameters entering this state. In a full agreement with theoretical predictions, entangled states with appropriate phase parameters can block transfer of excitations between different subsystems of a single composite quantum system (destructive quantum interference). This phenomenon can be interpreted in terms of bright and dark states known from quantum optics. Our results thus provide unambiguous demonstration for realization of entanglement and quantum interference effects in our modeling based on the real quantum device.

We also pointed out that the main sources of experimental inaccuracies in our modeling were errors in CNOT gates as well as  decoherence processes in the device. Attempts to implement two-step Trotter expansion led to the dramatic reduction of sensitivity of our results to phase parameters. This fact shows certain limitations in capabilities of current quantum computers.

Although the reported results can be relatively easily found explicitly or using classical computers, scaling towards chips with tens of physical qubits, improved coherence times and reduced CNOT errors might lead to the resolution of problems which can hardly be solved using more traditional approaches. Indeed, in order to study the dynamics from first principles, one needs to know all eigenstates of a given Hamiltonian. The number of eigenstates, in general, increases exponentially with the increase of particle number. Even for integrable systems (for instance, with Bethe ansatz techniques) the problem is formidable, since each eigenstate is characterized by its own solution of the set of Bethe equations. Finding numerically even single solution can be difficult. Moreover, even if solution is known, computation of overlap between corresponding Bethe vector and the initial state is not so easy. Therefore, even quantum computers of medium sizes, which can appear in the near future, might indeed be of practical importance for the modeling of dynamics of quantum systems.

\begin{acknowledgments}
 W. V. P. acknowledges a support from RFBR (project no. 15-02-02128). Yu. E. L. acknowledges a support from RFBR (project no. 17-02-01134).
\end{acknowledgments}

\appendix

\section{Preparation of 3PES}

\begin{figure}[h]
\includegraphics[width=1.00\linewidth]{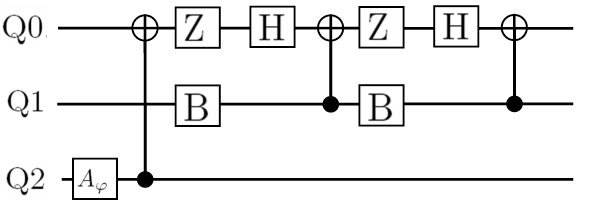}
\caption{
\label{3QES}
Quantum circuit for the preparation of three-qubit excited state.}
\end{figure}

Quantum circuit used to prepare 3PES is shown in Fig. \ref{3QES}. Single-qubit gates $A_{\varphi}$ and $B$ are constructed from the standard IBMqx4 gate $U_3$ as $A_{\varphi}=U_3(\theta = 2 \arccos \frac{1}{\sqrt{3}}, \varphi, \lambda = 0)$, $B=U_3(\theta = \frac{\pi}{4}, \varphi=0, \lambda = 0)$; Z is Pauli-Z gate.

\section{Full quantum circuits}

\begin{figure}[h]
\includegraphics[width=1.0\linewidth]{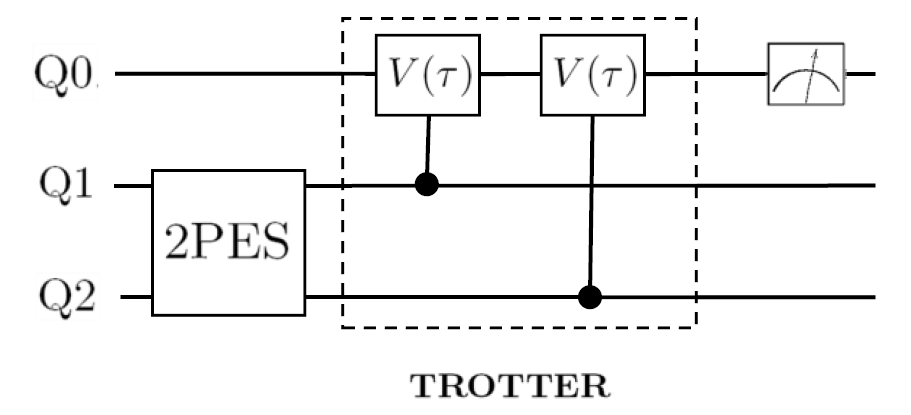}
\caption{
\label{fullcircuit1}
Quantum circuit for the evolution of the system starting from the initial state of two-particle entangled state of the bath and unexcited central spin.}
\end{figure}

\begin{figure}[h]
\includegraphics[width=1.0\linewidth]{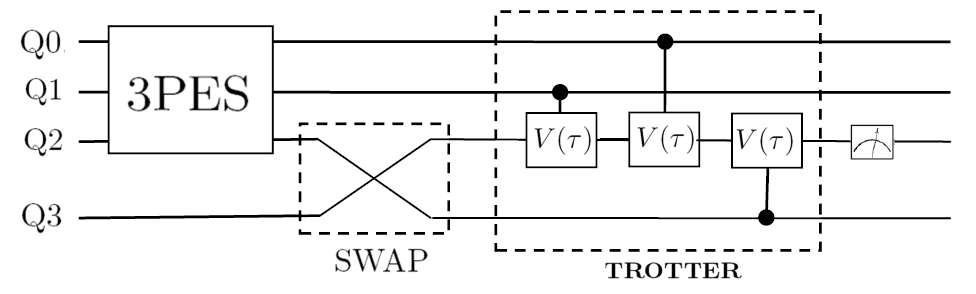}
\caption{
\label{fullcircuit2}
Quantum circuit for the evolution of the system starting from the initial state of three-particle entangled state of the bath and unexcited central spin.}
\end{figure}

\begin{figure}[h]
\includegraphics[width=1.0\linewidth]{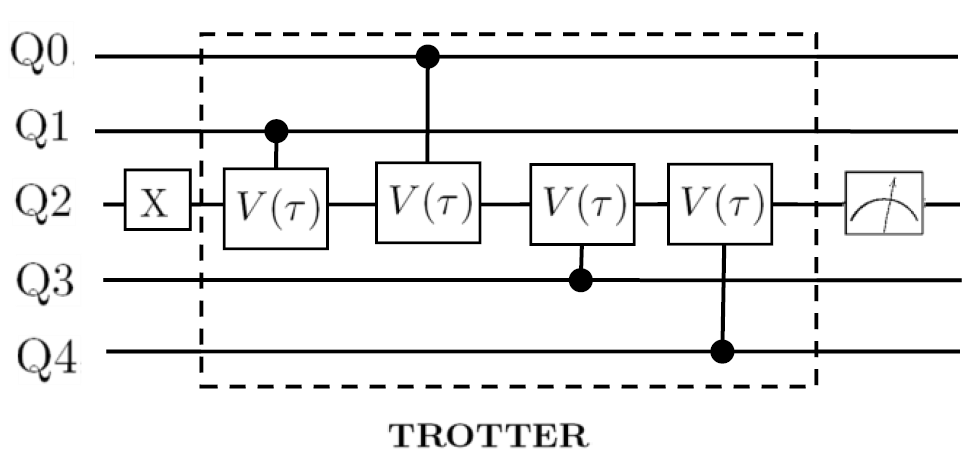}
\caption{
\label{fullcircuit3}
Quantum circuit for the evolution of the system starting from the initial state of excited central spin and four unexcited spins of the bath.}
\end{figure}

Figures \ref{fullcircuit1}, \ref{fullcircuit2}, and \ref{fullcircuit3} show full quantum circuits for three different systems and initial conditions we study.

\end{document}